# Pressure Sensor with Novel Electrical Circuit Utilizing Bipolar Junction Transistor

Mikhail Basov
Dukhov Automatics Research Institute, Moscow, Russian Federation
engineerbasovm@gmail.com

*Abstract*— **High sensitivity MEMS pressure sensor chip for different ranges (1 to 60 kPa) utilizing the novel electrical circuit of piezosensitive differential amplifier with negative feedback loop (PDA-NFL) is developed. Pressure sensor chip PDA-NFL utilizes two bipolar-junction transistors (BJT) with vertical n-p-n type structure (V-NPN) and eight piezoresistors (p–type). Both theoretical model of sensor response to pressure and temperature and experimental data are presented. Data confirms the applicability of theoretical model. Introduction of the amplifier allows for decreasing chip size while keeping the same sensitivity as a chip with classic Wheatstone bridge circuit.**

*Keywords—pressure sensor, bipolar junction transistor, differential amplifier, high sensitivity, temperature compensation.*

## I. Introduction

Piezoresistive pressure sensors are dominating in the pressure sensing market (automotive, medical, process control nuclear energy, aviation & space) and attracting a lot of R&D resources. The development of MEMS pressure sensors follows well-known trends in microelectronics, including chip size reduction while maintaining or improving sensor performance. The chip size reduction allows for chip and sensor cost reduction. Most of piezoresistive sensors use Wheatstone bridge electrical circuit with p-type resistors (PRs). Increase of sensor sensitivity while keeping the same chip size is one of the top priorities in sensor development. Researchers tried to solve this task by changing chip mechanical structure and PRs arrangement [1-19]. There are many attempts to combine wet etching (by KOH or TMAH) or/and deep reactive ion etching (DRIE) to create novel diaphragm structures. There are examples of significant size reduction of pressure sensor chips manufactured in medium and high volumes [20,21]. Sensing elements with profiled diaphragms can have high pressure sensitivity and linearity combined with small temperature errors. For example, sensing elements with very high pressure sensitivity demonstrated by Xi'an Jiaotong University [22,23]. However, sensitivity increase and chip size decrease by reducing membrane size and thickness has both process and design limitations. Further size reduction of pressure sensing elements can be achieved by adding active elements – transistors – and increasing the number of PRs in the electrical circuit. This work presents a pressure sensor chip (Fig. 1) with novel electrical circuit utilizing piezosensitive differential amplifier on bipolar junction transistors (BJT) with negative feedback loop (PDA-NFL) [24-29]. Circuit design and microstructure design allows for both significant increase of the sensitivity and for compensation of temperature errors. Performance of the PDA-NFL pressure sensing element is compared to other pressure sensing elements for three differential pressure ranges (1 kPa, 5 kPa and 60 kPa).

This work was supported by the Dukhov All-Russia Research Institute of Automatics (VNIIA).

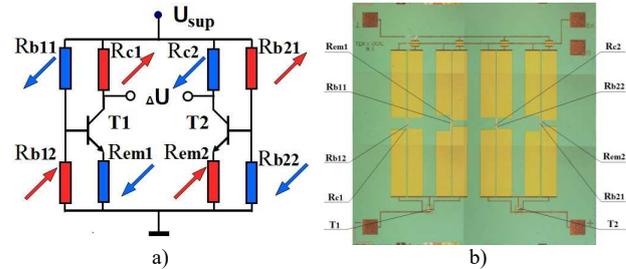

Fig. 1. Pressure sensor chip PDA-NFL: a) electrical circuit, b) top view.

## II. Theoretical and Technological Aspects

PDA-NFL pressure sensor chip was designed using both results obtained from an analytical mathematical model and from software modeling. Both lateral p-n-p type (L-PNP) and vertical n-p-n type (V-NPN) BJTs located on the membrane and on the frame (sensitive / not sensitive to pressure) were analyzed and included in the designs. Electrical circuit with NPN transistors used in the sensor chip is shown in Fig. 1a. Values of PRs and operating point of BJTs providing good balance between sensitivity and temperature compensation were determined analytically. Sentaurus TCAD software was used to finalize technological details. A process flow where all PRs, base of V-NPN BJT, emitter and collector of L-PNP BJT are created in one process step by diffusion of boron (p⁻) was chosen. The p⁻ layer has surface concentration $N_S = 6 \cdot 10^{18}$ cm$^{-3}$ (sheet resistance 200 Ohm/sq., p-n junction depth 2.2 μm). This allows for making PRs with high sensitivity to stress (piezoresistive coefficient $\pi_{44} = 1{,}26 \cdot 10^{-9}$ Pa$^{-1}$ for [110] direction). Phosphorous doping (n⁺) and drive-in were selected based on the target parameters of transistors. V-NPN BJT was designed to have gain $\beta_{V\text{-}NPN} = 150$, base current $I_{b\ V\text{-}NPN} = 5$ μA and collector-base voltage drop of 0.80 V. L-PNP BJT had target gain $\beta_{L\text{-}PNP} = 5$, base current $I_{b\ L\text{-}PNP} = 50$ μA; and collector-base voltage drop of 0.86 V (Fig. 2). All parameters of electrical circuit PDA-NFL presented in Table I. Four pairs of PRs placed in the areas with high mechanical stress (MS) selected with help of modeling in ANSYS (Fig. 3b). The target values of PRs were obtained by changing their length, while width of all eight PS was chosen to be 10 μm. P⁻ regions of PRs formed on the thin part of the membrane connected to metal lines on the frame by P⁺ areas. The chip uses Al-Si (1.5% Si) metallization. The PR resistances change as shown on Fig. 1a in response to applied pressure.

Pressure sensor chips were manufactured using (100) p-type Si wafers with n-type epitaxial layer. Required isolation of the BJTs was achieved by deep diffusion of p⁺ regions. The dimensions of pressure sensor chip with PDA-NFL are 4.0×4.0×0.4 mm³. It has square membrane with three rigid islands (RIs) formed by anisotropic wet etching in KOH (Fig. 3a and Table II). Sensors for three different pressure ranges: 60 kPa, 5 kPa and 1 kPa have been manufactured.

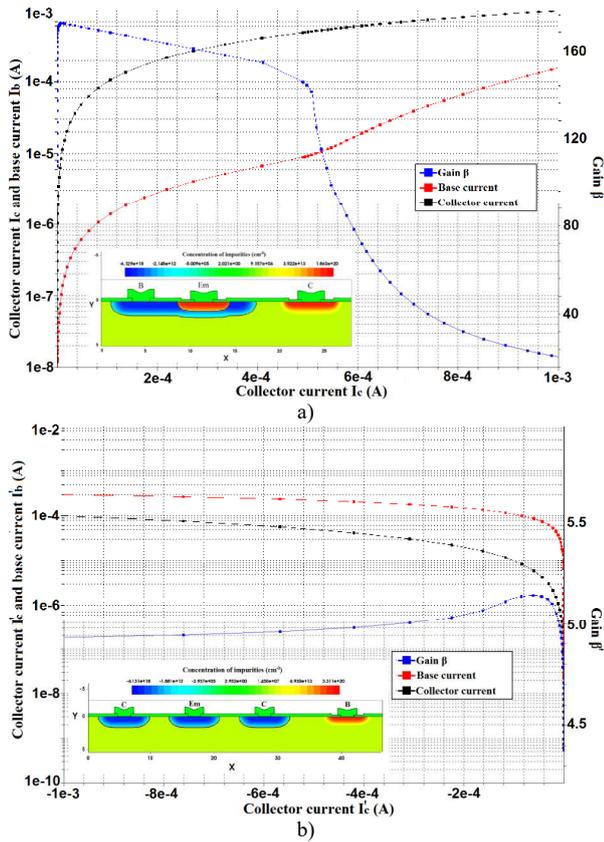

Fig. 2. Dependence of gain β on collector current (by TCAD software) for: a) V-NPN BJT, b) L-PNP BJT.

TABLE I. PARAMETERS OF PDA-NFL CIRCUIT.

| Parameters | V-NPN circuit | | L-PNP circuit | |
|---|---|---|---|---|
| | Analytical | Multisim | Analytical | Multisim |
| Base current BJT $I_b$, μA | 5.0 | 4.6 | 50.0 | 44.0 |
| Gain BJT β | 150 | 145 | 5 | 5 |
| Base-emitter voltage drop BJT ($U_b - U_{em}$), V | 0.80 | 0.80 | 0.78 | 0.77 |
| Collector-base voltage drop BJT ($U_c - U_b$), V | 0.81 | 0.80 | 0.86 | 0.86 |
| Collector potential BJT $U_c$, V | 2.50 | 2.79 | 2.50 | 2.81 |
| Parasitic current to substrate BJT $I_{par}$, μA | - | | 33.0 | |
| $R_{b11, B21}$, kOhm | 4.47 | | 3.00 | |
| $R_{b12, B22}$, kOhm | 2.98 | | 2.00 | |
| $R_{c1, c2}$, kOhm | 3.33 | | 10.00 | |
| $R_{em1, em2}$, kOhm | 1.79 | | 1.50 | |

TABLE II. GEOMETRICAL PARAMETERS OF PRESSURE SENSOR CHIP PDA-NFL FOR 60, 5 AND 1 kPa.

| Geometrical Parameter (Fig. 3a) | Size, μm | | |
|---|---|---|---|
| Pressure sensor chip PDA-NFL | 60 kPa | 5 kPa | 1 kPa |
| L | 4000 ± 20 | | |
| H | 400 ± 5 | | |
| A | 2260 ± 20 | | |
| W | 33 ± 2 | 9 ± 2 | |
| D | 41 ± 5 | 23 ± 5 | 39 ± 5 |
| Z | 490 ± 50 | 790 ± 50 | 270 ± 50 |

NI Multisim software was used for evaluation of chip sensitivity and additional temperature error based on the design parameters. Some results are shown in Fig. 4. Comparison of pressure sensitivity of the circuits with BJT sensitive / not sensitive to pressure shows that stress-sensitive BJTs placed on the membrane can increase pressure sensitivity only by about 1-5 % [30-32].

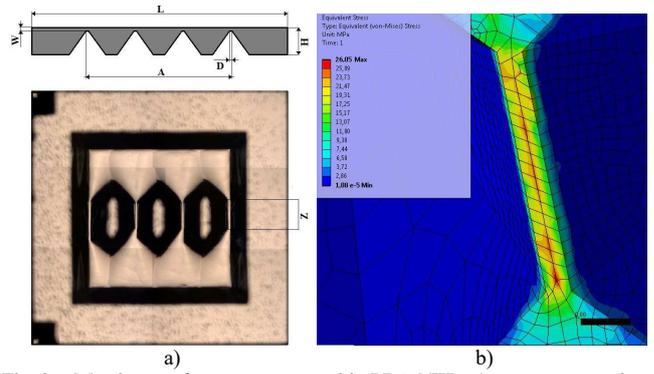

Fig. 3. Membrane of pressure sensor chip PDA-NFL: a) structure overview, b) stress distribution (von Mises) between RI by ANSYS in 1 kPa chip.

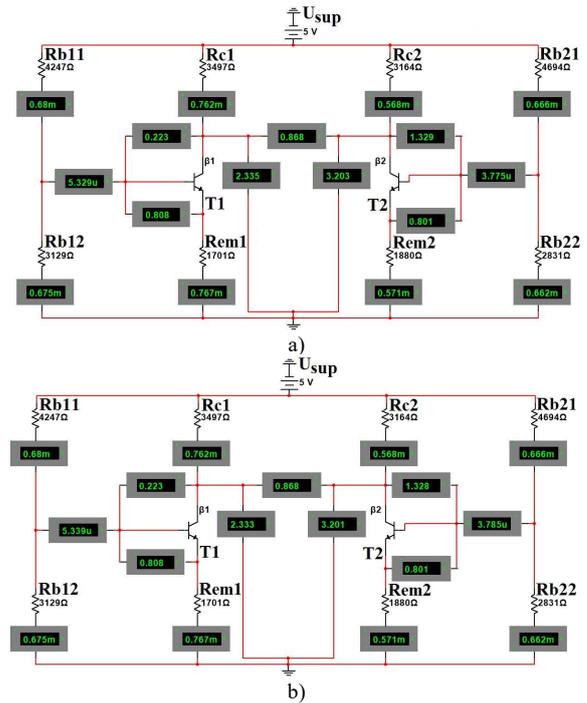

Fig. 4. Parameters of PDA-NFL circuits with non-deformable BJT V-NPN and for 60 kPa range calculated by NI Multisim on a) pressure ΔP = 100 kPa, b) temperature (parameters of left circuit branch has T = 37 ºC, parameters of right circuit branch – T = 27 ºC).

Placing BJTs on the frame has some advantages: it allows for achieving more symmetrical layout of components and avoiding connections between active and passive components on the membrane. Pressure sensor chip PDA-NFL with BJTs located on the membrane (pressure-sensitive) was fabricated but did not function properly due to a mistake in layout.

Theoretical model shows that electrical circuit PDA-NFL with V-NPN BJT is better than electrical circuit PDA-NFL with L-PNP BJT, for example for 60 kPa range: sensitivity $S_{V-NPN}$ = 1.736 mV/V/kPa and $S_{L-PNP}$ = 1.402 mV/V/kPa, additional temperature error for zero signal $TCZ_{V-NPN}$ = 0.026 %FS/°C and $TCZ_{L-PNP}$ = 0.584 %FS/°C. Parasitic current from emitter to substrate should be taken into account in the circuit utilizing L-PNP BJT (see Table II). This current exists even when p-n junction between epitaxial layer and substrate is closed. Analysis by TCAD shows that 67% of transistor current goes to the substrate. This fact has been confirmed experimentally. Therefore, section III contains only test data for pressure sensor chip PDA-NFL with V-NPN BJT located on the frame and not sensitive to pressure.

## III. EXPERIMENTAL DATA

Pressure sensor chip with PDA-NFL was bonded to a silicon support and placed on Kovar case (Fig. 5) for testing of its characteristics. The silicon support consists of an intermediate element and a pedestal, which are connected by low-temperature glass. The sensor assembly allows for measuring differential pressure. All samples were exposed to temperature and pressure cycling to remove residual assembly stress before functional testing [33-37].

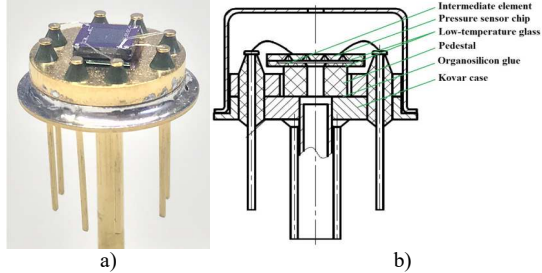

Fig. 5. (a) Pressure sensor PDA-NFL without cap, (b) assembly drawing.

Pressure sensors with PDA-NFL chip were compared with similar products utilizing Wheatstone bridge circuit having the same or similar pressure range: 1) mass-fabricated 60 kPa pressure sensor IPD60 (VNIIA), 2) high-sensitivity 5 kPa CBMP sensor (in development) [38,39], 3) ultra-high sensitivity pressure sensor [22,23] mentioned in Introduction. This sensor was developed for 0.5 kPa pressure range and its samples demonstrated sensitivity up to 66.0 mV/V/kPa. Data for 6 types of pressure sensors presented in Table III.

As it can be seen from Table III, sensors built with PDA-NFL chip have high sensitivity. In the 60 kPa range, pressure sensor with PDA-NFL chip has 3.65 times higher sensitivity than IPD60 sensor although both sensors have the same membrane size. Pressure sensor with PDA-NFL chip for 5 kPa has 2.2 times higher sensitivity, 7-8 times smaller hysteresis and improved repeatability compared to the CBMP pressure sensor even though CBMP sensor has 65% larger diaphragm area. Pressure sensor for 1 kPa with PDA-NFL chip has lower sensitivity than the chip with peninsula-island [22,23]. Yet, if both chips would have the same diaphragm size, then sensitivity of PDA-NFL chip (proportional to squared membrane length) will be about 109.6 mV/V/kPa or 66% higher. As PDA-NFL chip uses signal amplification, its mechanical structure is stronger and burst pressure is 2.5-5 times higher compared to the sensors with Wheatstone bridge [40].

Improvement of parameters of sensors with PDA-NFL chip is needed two areas. (1) The sensors have higher noise due to 1/f noise of BJTs. The noise has been minimized by optimizing manufacturing process and reducing an active base thickness of BJT [41]. Output noise also can be reduced by averaging multiple measurements in sensors with ASIC. (2) PDA-NFL circuit contains eight PRs and high temperature sensitive BJTs resulting in higher zero output signal and 2-3 times higher temperature errors. These parameters can be improved through PR layout optimization and optimization of temperature characteristics of BJTs.

## IV. CONCLUSION

The presented project demonstrates the opportunity for improvement of piezoresistive pressure sensors by on-chip integration of piezosensitive differential amplifier on BJTs. Theoretical model for pressure sensor PDA-NFL was created using both analytical calculations and software modeling. High-sensitivity sensing elements for pressure ranges (60 kPa, 5 kPa and 1 kPa) have been fabricated and tested. Pressure sensor with PDA-NFL showed significant advantages compared to piezoresistive pressure sensors for similar pressure ranges with Wheatstone bridge circuit. Directions for further improvement of pressure sensors with PDA-NFL electrical circuit on the chip are discussed.

TABLE III. OUTPUT CHARACTERISTICS OF PRESSURE SENSORS CHIPS: PDA-NFL, IPD60, CBMP AND PENINSULA-ISLAND.

| Type of pressure sensor chip | | PDA-NFL | IPD60 | PDA-NFL | CBMP [38,39] | PDA-NFL | Peninsula-island [22, 23] |
|---|---|---|---|---|---|---|---|
| Pressure range | | 60 kPa | | 5 kPa | | 1 kPa | |
| Dimension of membrane thin part, mm² | | 2.26×2.26 | | 2.26×2.26 | 2.90×2.90 | 2.26×2.26 | 3.50×3.50 |
| Number of samples in statistics | | 20 | 1200 | 22 | - | 14 | - |
| Sensitivity S, mV/V/kPa | | 1.877 | 0.515 | 11.24 | 5.14 | 44.9 | 66.0 |
| Zero pressure output signal (Offset), mV/V | | 6.8 | 3.0 | 8.0 | 1.3 | 14.0 | - |
| Noise voltage, µV/V | | 3 | 0.8 | 3 | - | 12 | - |
| Nonlinearity 2$K_{NL}$, %FS | | 0.059 | 0.055 | 0.27 | 0.28 | 0.26 (for 0.5 kPa) | 0.33 (for 0.5 kPa) |
| Hysteresis H, %FS | | - | - | 0.03 | 0.26 | 0.28 (for 0.5 kPa) | 0.36 (for 0.5 kPa) |
| Repeatability R, %FS | | - | - | 0.08 | 0.53 | 0.42 (for 0.5 kPa) | 0.67 (for 0.5 kPa) |
| Zero change after flipping sensor in gravity field, µV/V/g | | - | - | 11 | - | 13 | 28 |
| Burst pressure $P_{burst}$, kPa | | 1600 | | 600 | - | 500 | 105 |
| Output changing after pressure overload | of offset, %FS | 0.006 | 0.004 | 0.023 | - | 0.083 | - |
| | of pressure sensitivity, %FS | 0.069 | 0.072 | 0.081 | - | 0.061 | - |
| Long-term stability | of offset, %FS | 0.007 | 0.002 | 0.032 | - | 0.122 | 0.138 |
| | of pressure sensitivity, %FS | 0.006 | 0.003 | 0.023 | - | 0.166 | - |
| Thermal hysteresis of zero, %FS | (−30…+20°C) | 0.065 | 0.033 | 0.042 | - | 0.341 | - |
| | (+20…+60°C) | 0.045 | 0.025 | 0.050 | - | 0.262 | - |
| Thermal hysteresis of span, %FS | (−30…+20°C) | 0.042 | 0.018 | 0.189 | - | 0.611 | - |
| | (+20…+60°C) | 0.017 | 0.012 | 0.061 | - | 0.343 | - |
| Thermal coefficient of zero, (%/°C)FS | (−30…+20°C) | 0.007 | 0.003 | 0.014 | - | 0.092 | 0.165 |
| | (+20…+60°C) | 0.010 | 0.002 | 0.012 | 0.013 | 0.096 | |
| Thermal coefficient of span, (%/°C)FS | (−30…+20°C) | 0.020 | 0.043 | 0.222 | - | 0.633 | - |
| | (+20…+60°C) | 0.045 | 0.034 | 0.204 | 0.116 | 0.634 | - |